# Research on work zone vehicle queuing behavior based on cellular automata

ShaoYuan.CHEN[a], LiXian.ZHONG[a], RuiXi.ZHU[b], LianSheng.YANG[a], Mirizhati AIKEBAIER[a], CiYun.LIN[a,c]*

a College of Transportation, Jilin University, Changchun 130022, China
b College of Communications Engineering, Jilin University, Changchun130022, China
c State Key Laboratory of Automobile Simulation and Control, Jilin University, Changchun 130022, China
* Corresponding author. Email address: linciyun@jlu.edu.cn

**Abstract**

A model is proposed to estimate the work zone queue length, and the cellular automata based on empirical data is used for model validation. This estimation model can be applied to work zone organization and management to improve work zone capacity and security. Relationship between the average queue length and the warning zone length can be found, and the appropriate warning zone length can be determined according to design flow. Moreover, the appropriate work zone lane-changing strategies under different design flows are found through the estimation model.

## 1 Introduction

Due to infrastructure construction and reconstruction on the highway, it can be seen that the work zone occupies the highway space and forms a traffic bottleneck. The work zone results in capacity reduction and delay growth [ 1-3 ]. At the same time, vehicles in the closed lane are forced to change lane, which increases the conflicts, resulting in the increase of the possibility of crashes and the reduction of driving security [ 4-6 ].

Vehicles in the closed lane must find the appropriate gap to merge into the open lane. With the increase of traffic volume, it is possible that part of vehicles in the closed lane stay in front of the work zone. First, the delay time of this part of vehicles can be significantly extended. Second, these vehicles may change lane aggressively and increase the risk of crashes. Third, the remaining vehicles occupy the highway space, which is equivalent to extending the length of the work zone. Therefore, it is attached to great significant to estimate the work zone queue length.

Researchers have done a lot of research on the traffic flow characteristics at work zone, and different models have been established. Weng and Meng ( 2015 ) [7] established the speed-flow relationship and capacity model for work zone, and verified the influence of work zone configuration conditions on speed-flow relationship and capacity of work zone. Weng et al. ( 2018 ) [8]established a mixed time-varying logit model for the vehicle merge behavior in the pre-merge area of the work zone. Duan et al. ( 2020 ) [9]established a survival model to analyze the influence of various factors of environment and driver 's personal characteristics on the decision-making distance and parallel distance in the lane change process, and improved the research on other factors affecting the merge of work zones. In the past, some studies have focused on the evaluation of queuing length in the work zone. Ramezani et al. ( 2010 ) [10]studied the queue length prediction of stranded vehicles in the work zone. Ramezani et al. ( 2011 ) [11]proposed two bottleneck locations in the work zone, namely the work zone and the transition area, and predicted the queue length in this scenario. Ramezani et al. ( 2012 ) [12] established a more accurate model to predict the queue length assuming that there were two bottleneck locations in the work zone. Venter, L and Bester, CJ ( 2015 ) [13]studied the work zone with stop / line control and predicted the back-of-queue position at such a work zone. Based on a large number of measured data, Gilandeh, MM ( 2021 ) [14]proposed a model for predicting the queuing length for expressway work zone. However, in the past studies, the calculation is usually based on the traffic wave theory, which lacks consideration of the psychological factors that promote the driver to change lane.

VISSIM and CA, two common simulation methods, are used to simulate work zone traffic. Park et al. ( 2013 ) [15]used VISSIM to simulate the selection of variable speed limit strategy ( VSL ) and the installation interval of portable information signboard ( PCM ) under different conditions in the work zone. Huang and Wang ( 2015 ) [16]used VISSIM to simulate the traffic capacity of the work zone, and gave suggestions for speed limit. Zhang et al. ( 2016 ) [17]calibrates the expected velocity distribution of vehicles in the work zone by field survey data, and simulates the expected velocity curve in VISSIM to verify the effectiveness of the calibration method. However, VISSIM has poor simulation effect on abnormal traffic flow under special conditions such as work zone and traffic accidents, and its simulation results cannot well reflect the real traffic situation under the influence of work zone. On the contrary, cellular automata is more suitable for abnormal traffic flow.

Meng and Weng ( 2010 )[18] regards randomization probability as a function of flow rate and work zone configuration conditions, and establishes cellular automata for simulating work zone. Later, Meng and Weng ( 2011 )[19] proposed an improved cellular automata, which added new rules of horizontal velocity and position update to better reflect the traffic in the work zone. Hou and Chen ( 2019 )[20] developed an improved cellular automaton to compensate for the shortcomings of previous cellular automaton models in driving style and deceleration behavior. Cellular automata are widely used in the study of abnormal traffic flow under the influence of construction. Xiao, W et al. ( 2017 ) [21]studied the traffic behavior of intelligent two-way system with work zone by using cellular automata. Das and Chattaraj ( 2019 )[22] established a cellular automaton to simulate the traffic in the work zone, and obtained the negative effects due to lane drop at work zone. Wenjing Wu et al. ( 2020 )[23] used cellular automata to simulate human-driven vehicles and intelligent networked autonomous vehicles ( CAV ), and simulated and evaluated the speed and lane change control strategy of intelligent networked autonomous vehicles ( CAV-based ) in expressway work zone under heterogeneous traffic flows.

There are also a lot of research on traffic organization optimization of work zone. Some lane-changing strategies have been proposed in the past. The main content of the normal merge ( NM ) strategy is to set up the warning zone and install the signboard, which indicates that there is a work zone in front of the drivers. When the traffic volume increases, the warning zone cannot be extended indefinitely, and normal merge strategy can not meet the traffic demand. Therefore, Nemeth et al. ( 1982 )[24] proposed early merge ( EM ) strategy to promote closed-lane drivers to change lane farther from the work zone. In order to improve the utilization efficiency of closed lanes, the late merge ( LM ) strategy was proposed. Walters et al. ( 2000 )[25] verified the effectiveness of the late merge strategy. Both early and late merges can increase the merge behavior in the closed lane and reduce congestion. However, when the traffic flow increases to a certain threshold, the headway in the open lane is small, and the number of lane changeable gaps reduces. Both early and late merges are difficult to adapt to this situation. Therefore, Yang et al. ( 2009 )[26] proposed a lane-by-lane signal control method to split the traffic flow of different lanes in time. And achieved good effect of easing congestion. Duan et al. ( 2022 )[27] proposed the application of variable signal phase control in the work zone, and compared the results of this strategy with those of static early merge, static late merge, normal merge and fixed signal phase control. However, in the past studies, when discussing the comparison and selection of lane change strategies in the work zone, there was a lack of attention to the microscopic parameters such as queue length. In addition, although the warning zone length division is actually important for construction organization optimization, the research on the warning zone length is still lacking [28].

This study makes up for the deficiency of the existing research mentioned above. This study establishes a estimation model for the average queue length of the work zone: the queue length generated by the closed lane of the work zone in unit time can be estimated by the current or historical traffic data, that is, the number of retained vehicles increased in the closed lane of the work zone in unit time. The model takes into account two different driving forces that psychologically promote the driver to change lane on closed lanes, namely objective stimuli and value stimuli, and calibrates the weight of objective stimuli and value stimuli in driver ' s lane change decision-making. The meaning of objective stimuli and value stimuli will be explained in the part of model establishment. The idea and process of the model established in this study have strong innovation, and provide a good idea for the study of traffic flow characteristics with traffic bottlenecks. On the model validation, this study uses the cellular automaton traffic flow model based on empirical data. After the verification of the model, this study uses the estimation model of the average queue length of the work zone to quantitatively study the calibration of the length of the warning zone, and takes the microscopic parameter of the average queue length of the work zone as the evaluation index to discuss the selection of the lane change strategy in the work zone, and finally obtains the best selection results of the lane change strategy under different design flow.

This paper involves model establishment, model validation and model application. Firstly, the ' supply and demand relationship ' of the lane-changing demand in the closed lane and lane-changeable gaps in the open lane is analyzed and the estimation model of the average queue length of the work zone is established. Secondly, the cellular automata are used to verify the estimation model. Finally, the estimation model is applied to the calculation of the appropriate warning zone length as well as the comparison and selection of work zone merge strategies.

**2 Model establishment**

In this section, the 2to1 two-lane highway work zone is analyzed as the research object. Through this model, the average queue length can be estimated with the current or historical data of traffic volume. Work zone area is divided into work zone, warning zone and normal driving zone. There is no warning sign in the normal driving zone. The upstream of the work zone is demonstrated in Figure 1.

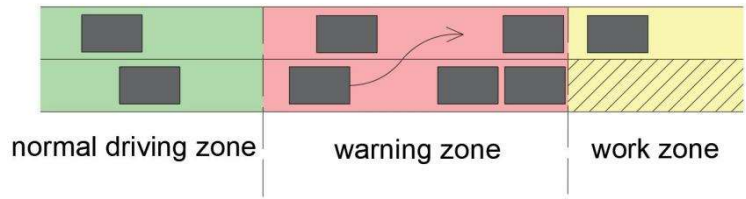

Figure 1 The upstream of the work zone

The basic assumptions of the model are as follows:

（1）The flow entering the road section from the upper limit of the normal driving area is divided equally by two lanes, assuming that the arrival rate of a single lane is $\lambda$；

（2）The arrival rate of each lane in the upper limit of the warning zone is equal to that of the corresponding lane in the upper limit of the normal driving area；

（3）The implementation of normal merge（NM）strategy, that is, only warning signs and guardrails are set up, and no additional construction organization scheme is adopted；

（4）Open lane vehicles will not merge into the closed lane；

（5）The arrival of vehicles in the upper limit of the warning zone obeys Poisson distribution.

Vehicles in the closed lane must change lane from the original lane to the open lane to bypass the work zone. It is assumed that during a certain period of time, the maximum number of vehicles allowed to merge into the open lane is the total lane-changing supply S, and the number of vehicles required to change lane in the closed lane is the total lane-changing demand D. Theoretically, the number of queue vehicles in the closed lane is equal to max（D-S, 0）. If S > D, it means all vehicles in the closed lane have changed lane. However, in fact, even if S > D, it is still likely to see congestion. The reason is that the lane-changing supply of the open lane and the lane-changing demand of the closed lane are unevenly distributed in the warning zone. Where the location is closer to the work zone, the gaps for lane-changing in the open lane is reduced, while the driver's psychological demand for lane-changing is increasing as vehicles get close to the work zone. This imbalance leads to the waste of front section clearance and the shortage of rear section clearance, and the work zone is prone to produce stranded vehicles.

The upper limit of the warning zone is zero, the road section is X axis, the direction to the work zone is X axis positive direction, x represents the position of the vehicle on the X axis, and the arrival rate in the closed lane at x position is $\lambda(x)$. In the study of lane-changing behavior, it is assumed that the driver in the closed lane first decides whether to change lane, if not, to continue along the original lane, if so, to judge whether there is a lane-changeable gap in the open lane, if not, to continue along the original lane, if so, to change lane. At each location, there is a lane-changing decision. If the driver does not think and judge, and continue in the original lane out of psychological inertia, this case is still regarded as a decision, and the decision result is not to change lane.

This paper uses s（x）and d（x）to describe the lane change supply of open lanes and the lane change demand of closed lanes. The greater the s（x）, the greater the lane change supply of the open lane at x position. Similarly, the greater the d（x）, the greater the lane change demand of the closed lane at x position. Figure 2 shows 'supply and demand relationship'.

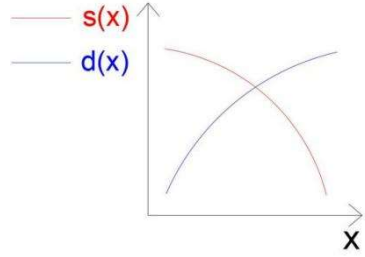

Figure 2 Diagram of Supply and Demand Relationship

The lane-changing supply s ( x ) is actually the possibility of a lane-changeable gap at the x location, and the value range is [ 0,1 ]. It can also be understood that a longer period of time is divided into several time intervals, each time interval can make a vehicle pass through this location at the design speed of the road section. The time interval is numbered in chronological order i = 1,2, 3..., n. If there is a lane-changeable gap in the open lane within the time interval i, then T ( i ) = 1, and vice versa, then T ( i ) = 0. Finally, the expected value of T ( i ) ( i = 1,2,3,..., n ) ,E(T) is calculated. E(T) is the lane change supply s ( x ) at x. At position x, the arrival rate in the closed lane is $\lambda$ ( x ), and the vehicle arrival in the open lane is $2\lambda - \lambda(x)$. Assuming that the minimum lane-changeable gap is $t_c(s)$, and the arrival rate is Poisson distribution, $s(x) = e^{-[2\lambda-\lambda(x)]t_c}$. The minimum lane-changeable gap $t_c$: when the lane-changing vehicle enter the new lane, it can just find a position, and maintain the minimum safety distance with the front and rear vehicles. Assuming that the minimum safety headway is $h_0$ and the design speed of the vehicle is $v_0$, the minimum lane-changeable gap is $t_c = 2h_0/v_0$. Ghasemi, SH et al. ( 2016 )[29] studied the minimum safe headway of highway vehicles, and pointed out that the average minimum safe headway was between 145 and 165 feet. Here, take 145 feet, convert 1ft = 30.48cm, get $h_0$= 44.196m. Taking into account the constraints of the work zone, when discussing the value of the design speed, the service level of the expressway is taken as the level of E service, and the vehicle speed generally does not exceed 80 km / h ( HCM )[30]. Here, the design speed is taken as the upper limit, namely 80 km / h. Finally, $t_c = 2h_0/v_0 = 3.97764s \approx 4s$.

The lane-changing demand d( x ) is the possibility of the driver choosing the lane-changing at position x when there is a lane-changing gap, and the value range is [ 0, 1 ], reflecting the psychological needs of the driver for lane-changing. Assuming that the frequency of lane change completed by the driver at position x is n( x ), then d ( x ) = n ( x ) / [ s ( x ) * λ ( x ) ], where the product of lane change supply and vehicle arrival is actually the ' event ' : at position x, the frequency of both drivers passing through the simultaneous existence of lane change gap ', and the ratio of n ( x ) to it is the possibility of the driver choosing lane change in the case of lane change gap, namely, lane change demand d ( x ). Thus, ( x ) s ( x ) d ( x ). At the next position, the arrival of the vehicle in the closed lane becomes ( x + △ x ) = λ ( x ) − n ( x ) ( △ x is a small length interval ).

So, there are:

$$\begin{cases} s(x) = e^{-[2\lambda-\lambda(x)]t_c} \\ n(x) = \lambda(x)s(x)d(x) \\ \lambda(x+\triangle x) = \lambda(x) - n(x) \end{cases}$$

Boundary conditions: λ ( x = 0 ) = λ

If the expression of lane change demand can be obtained, the vehicle arrival at each location can be calculated by the above recursive formula. Next, the lane change demand d ( x ) is analyzed and calibrated.

The lane changing is classified into two types: mandatory lane changing and discretionary lane changing [31]. Mandatory lane changing is a forced lane-changing behaviour for a certain purpose, and discretionary lane changing is a voluntary lane-changing behaviour to pursue better driving conditions to improve driving speed ( or reduce travel time )[32]. Inspired by the classification of lane change, this study proposes the concepts of objective stimuli and value stimuli, and believes that for drivers in the closed lane, there are two kinds of driving forces ( stimuli ) that promote them to change lane: objective stimuli and value stimuli. Objective stimuli is caused by obstacles that force drivers to change lane, which may be caused by blockage, from the original lane with obstacles to the destination lane without obstacles. Value stimuli comes from the driver ' s pursuit of greater driving efficiency, which can be expressed as driving speed, headway, or their combination. These two driving forces occupy a certain weight when influencing the driver ' s decision-making for lane changing. During the decision-making of drivers with different driving styles, the weight of objective stimuli and value stimuli is different. For example, some drivers are less interested in greater driving benefits, and some drivers will make radical lane-changing behavior to pursue greater driving benefits. However, in a large sample, the driver set presents an overall ' average ' weight of objective stimuli and value stimuli, that is, α and 1-α referred to below, representing the weight of objective stimuli and value stimuli embodied in the driver set, respectively. The weight of the two driving forces is the inherent property of the driver set. The value of α is not related to traffic volume and the surrounding environment, which will be verified in the model verification section.

In summary, d ( x ) is represented as
$$d(x) = \alpha O + (1 - \alpha)V$$
$\alpha -$ the weight of objective stimuli, $\alpha \in [0,1]$
$O -$ the intensity of objecitive stimuli, $O \in [0,1]$
$V -$ the intensity of value stimuli, $V \in [0,1]$
The specific calculation methods of O and V are as follows.
Calculation of O:

In the work zone lane changing scene, the factors influencing the intensity of objective stimuli include the distance from the work zone L-x and the possibility of lane-changeable gap s ( x ). If only look at s ( x ), then s ( x ) and O roughly showed a negative correlation, that is, with the decrease of the gap, the intensity of objective stimuli will increase. The distance between the vehicle and the work zone is considered to affect the sensitivity of O to s ( x ) in this model. Specifically, in the position far from the work zone, the driver's purpose of lane-changing is weak. However, if the traffic flow density increases and the lane change gap decreases, the driver is more likely to consider lane-changing in advance, and O is significantly improved. However, in the position near the work zone, due to the proximity to the work zone, whether the traffic flow density in the open lane is large or small, the driver's lane change demand is relatively strong, and the increase or decrease of the lane change gap in the open lane has little effect on the purpose driving value. The larger the value is, the greater the sensitivity of O to s ( x ) is. A preliminary function relation is obtained
$$O = -(L - x)s(x) + C$$
C- a constant to be determined
L- Length of the warning zone.

It is not difficult to find that the scope of O function is not in [ 0,1 ], so it needs to be standardized so that O function is in [ 0,1 ].

$$O = \frac{-(L-x)}{L}s(x) + C$$

Because $s(x) = e^{-(2\lambda - \lambda(x))t_c}$, and $\lambda(x)$ is monotone decreasing function, the O function is monotone increasing function. Thus, C can be transformed into $Cx + D$, and it does not affect the overall trend of the function.

$$O = \frac{-(L-x)}{L}s(x) + Cx + D$$

- All parameters to be determined

According to the actual situation of the driver's psychology, two special solutions, namely $x = 0$, $O = 0$; $x = L$, $O = 1$.

$$O = \frac{-(L-x)}{L}s(x) + \frac{1 - e^{-\lambda t_c}}{L}x + e^{-\lambda t_c}$$

$$O = \left(\frac{x}{L} - 1\right)(s(x) - s(0)) + \frac{x}{L}$$

Calculation of V:

V is determined by the driving efficiency difference between the two lanes. Specifically, the driving efficiency difference can be shown as the difference in the space headway between the open lane and the closed lane, that is, the difference of $h_{si}(i = 1,2)$. $h_{s2}$ and $h_{s1}$ respectively represents the space headway of the open lane and the headway of the closed lane. However, the space headway does not take into account the vehicle speed of different lanes. Therefore, in this paper, the driving efficiency is represented by the time headway. According to the traffic flow theory, the basic parameters of traffic flow are expressed as $q = kv$, where, q is the traffic volume ( pcu / h ), k is the density ( pcu / km ), v is the speed ( m / s ). Here, q of a certain location is the arrival rate $\lambda(x)$ of this location. $q = \frac{3600}{\overline{h_t}}$, $k = \frac{1000}{\overline{h_s}}$, so $\frac{3.6\overline{h_s}}{v} = \overline{h_t}$. If the calculation time is relatively small, there is approximately, $\overline{h_t} = h_t$, $\overline{h_s} = h_s$. For each lane, $V_i = h_{ti}(i = 1,2)$ (representing the driving efficiency of closed-lane vehicles and open-lane vehicles respectively ). At the position x, $V_1 = h_{t1} = 3600/\lambda(x)$, $V_2 = h_{t2} = 3600/(2\lambda - \lambda(x))$

But at this point the scope of the V function is not in [ 0,1 ], need to be standardized. In this paper, LOGIT model is used for standardization.

$$V = \frac{e^{V_2}}{e^{V_1} + e^{V_2}}$$

To sum up, the calculation formula of another angle.

$$d(x) = \alpha\left[\left(\frac{x}{L} - 1\right)(s(x) - s(0)) + \frac{x}{L}\right] + (1 - \alpha)\left(\frac{e^{h_{t2}}}{e^{h_{t1}} + e^{h_{t2}}}\right)$$

Finally, the recursive model is obtained:

$$\begin{cases} s(x) = e^{-[2\lambda - \lambda(x)]t_c} \\ d(x) = \alpha\left[\left(\frac{x}{L} - 1\right)(s(x) - s(0)) + \frac{x}{L}\right] + (1 - \alpha)\left(\frac{e^{\frac{3600}{2\lambda - \lambda(x)}}}{e^{\frac{3600}{\lambda(x)}} + e^{\frac{3600}{2\lambda - \lambda(x)}}}\right) \\ n(x) = \lambda(x)s(x)d(x) \\ \lambda(x + \Delta x) = \lambda(x) - n(x) \end{cases}$$

Boundary conditions: λ(0) = λ

The values need to be determined by combining traffic flow simulation technology. In this study, cellular automata (CA) model is used to simulate the actual road conditions. According to a series of values, different approximate solutions are obtained. Finally, the value trend is judged according to the polynomial fitting results, and the value is obtained by combining the statistical unbiasedness method. The specific process will be described in detail in the model validation section.

**3 Model Verification**

3.1 Model Discretization

Although our model could realistically reflect the driver's lane change behavior and the relationship between transportation supply and transportation demand, it still needs scientific methods for verifying model and comparison between different strategies. Above purpose could be well achieved by combining discrete model with the Cellular Automata models.

In order to apply computer technology to our model better for comparison with the results of traffic simulation test and complete a series of work such as parameter calibration, it is necessary to discretize our model. The discrete model is similar to operation mode of the CA model, therefore accuracy of model will not be reduced due to discretization. The specific model discretization process is as follows

The continuous value of position coordinate $x$ is transformed into the discrete value $k$, any sequence $|x_k|$ satisfying infinite increase according to Heine Theorem, there are $\lim_{x \to x_o} f(x) = A$.

Therefore, the discretization of continuous position coordinate value x does not affect accuracy of each parameter of model. Accordingly, the discrete model is derived as follows

$$s(k) = e^{-[2\lambda - \lambda(k)]t_c}$$

$$d(k) = \alpha\left(\left(\frac{k}{L} - 1\right)(s(k) - s(0)) + \frac{k}{L}\right) + (1 - \alpha)\left(\frac{e^{\frac{3600}{2\lambda - \lambda(k)}}}{e^{\frac{3600}{\lambda(k)}} + e^{\frac{3600}{2\lambda - \lambda(k)}}}\right)$$

$$n(k) = \lambda(k)s(k)d(k)$$

$$\lambda(k+1) = \lambda(k) - n(k)$$

where $\lambda(0) = \lambda$, $k = 0,1,2 \ldots L$

The model of predicting the average queue length in the closed lane was established, but the value of weight coefficient α is still not calibrated. Considering main variable traffic conditions include arrival flow λ in our model, the value of α calibrated by determining the relationship between arrival flow λ and weight coefficient α in the model verification part, and explains the physical meaning of weight coefficient α. Finally, we carry out the error analysis for verifying the prediction accuracy of our model.

3.2 Cellular Automata Rules

Here we use Fei et al (2016) [33] has built a cellular automaton that has been validated by empirical data to simulate real traffic flows as a basis for model verification and parameter calibration. The principle of this cellular automata is as follows.

Vehicles are divided into two types, each accounting for a certain proportion: good performance type and poor performance type, which have different maximum speeds and accelerations. The acceleration is also different in high / low gear.

There are two types of drivers, each accounting for a certain proportion: cautious type and radical type, and there are different status update rules for different drivers.

Vehicle Generation:

Only generates new vehicles at (1, 1), (2, 1).

The conditions for generating a new vehicle are:
$$d_0(t) \geq L_e$$
$$\rho_{curr} < \rho$$

Among them:
$$d_0(t) = x_1(t) - x_0(t) - L_v$$
$$\rho_{curr} = N/(2L - L_w)$$

Where $d_0(t)$ represents the distance between the newly generated vehicle on the lane at time t and the vehicle in front, $x_0(t)$ is the abscissa of the newly generated vehicle on the lane at time t, $x_0(t) = 1$, and the vehicle length $L_v = 7$. If there is no front car, $d_0(t) = +\infty$. $\rho_{curr}$ represents the vehicle density in the current CA model, $\rho$ which is the specified vehicle density. N is the total number of vehicles, L is the length of the lane, and Lw is the length of the construction area.

Update the rule:

Accelerations:
$$v_n(t) \leftarrow \min(v_n(t) + a, v_{max})$$

Randomly slowed down with probability p:
$$v_n(t) \leftarrow \max(v_n(t) - 1, 0)$$

Decisive deceleration:
$$v_n(t+1) \leftarrow \min(v_n(t), d_n), \text{cautious drivers}$$
$$\begin{cases} v_{n+1}(t+1) \leftarrow \max(\min(v_n(t) + a, v_{max}(n+1), d_{n+1}(t)), 0) \\ v_n(t+1) \leftarrow \min(v_n(t), d_n(t) + v_{n+1}(t+1)) \end{cases}, \text{radical drivers}$$

Location Update:
$$x_n(t+1) \leftarrow x_n(t) + v_n(t+1)$$

Where $v_n(t)$ represents the speed of the nth vehicle cell from left in a lane in the CA model at time t, Vmax represents the maximum speed of this type of vehicle, $x_n(t)$ represents the abscissa of the vehicle at time t, $d_n(t)$ represents the distance between the vehicle and the vehicle in front at time t, which meets the requirements of $d_n(t) = x_{n+1}(t) - x_n(t) - L_v$ and vehicle length $L_v = 7$; if there is no forward vehicle, $d_n(t) = +\infty$.

Lane Change Rules:

Lane change in non-warning areas:

incentive criterion：
$$\begin{cases} d_n < \min(v_n + a, v_{max}) \\ d_{pred} > d_n \end{cases}$$

security criterion：
$$d_{succ} > \min(v_{succ} + a_{max}, v_{s,max}) - \min(v_n + a, v_{max})$$

To change lanes in the warning zone:

incentive criterion：
$$d_{pred} > L_e$$

security criterion：

$$d_{succ} \geq v_{s,max}, \text{cautious, drivers}$$
$$d_{succ} > \min(v_{succ} + a_{max}, v_{s,max}) - \min(v_n + a, v_{max}), \text{radical drivers}$$

Where, $d_{pred}$ represents the distance between the vehicle and the nearest vehicle in front of the target lane of lane change. If there is no vehicle in front, $d_{pred} = +\infty$. $d_{succ}$ indicates the distance between the vehicle and the nearest vehicle behind the lane where the lane is changed. If there is no vehicle behind, $d_{succ} = +\infty$. $L_e$ is the minimum forward distance, $L_e = 5$. $v_{s,max}$ represents the maximum speed of the nearest vehicle behind the lane on the target lane of lane change.

In addition, for the construction area cell, when it is used as the front vehicle of a vehicle, it is considered as a stationary vehicle with v = 0 and a = 0. When it is the rear vehicle of a vehicle, it is deemed that there is no vehicle behind the vehicle.

Simulation related parameters are shown in the Table 1 below.

Table 1 Relevant parameters of cellular automata simulation

|  | Warning zone | Work zone |
|---|---|---|
| Vehicle random acceleration probability pAcc | 1.1 | 1.1 |
| Vehicle random deceleration probability pSd | 0.2 | 0.2 |
| The probability of switching under the condition of lane change in the warning area is satisfied pChange | 0.7 | 0.7 |
| Ratio of different models typeDistribute | 0.3 | 0.3 |
| The proportion of radical drivers drivesRatio | 0.25 | 0.25 |
| Acceleration at startup startupAcc | 3 | 2 |
| Acceleration in motion movingAcc | 1 | 1 |
| Maximum vehicle speed maxV | 28 | 17 |
| Warning and construction area speed limit specialLimitV | 14 | |

3.3 Parameter of weight value α Calibration

By inputting the parameter arrival flow λ of the CA model, queue length in the closed lane is obtained. (Parameters of the CA model are calibrated as follows). At the same time, the arrival flow λ which is the same as the CA model one is inputted onto our model. By continuously adjusting the value of α -a step size is 0.01, the prediction queue length of closed lane and its corresponding α value are obtained, which are the closest to queue length in the CA model. The value of corresponding α is obtained by continuously adjusting the arrival flow λ, finally the different values of the arrival flow λ and the weight coefficient α are obtained as follows Table 2 and Figure 3.

Table 2 Different arrival flow and corresponding α values

| arrival flow (veh/h/lane) | 494.4 | 574.3 | 639.3 | 697.8 | 750.9 | 796.6 | 838.3 | 875.6 | 904 |
|---|---|---|---|---|---|---|---|---|---|
| alpha | 0.17 | 0.18 | 0.23 | 0.19 | 0.19 | 0.21 | 0.21 | 0.18 | 0.18 |
| arrival flow (veh/h/lane) | 906.2 | 906.5 | 906.5 | 906.9 | 907.7 | 908.6 | 908.8 | 909 | |
| alpha | 0.21 | 0.19 | 0.21 | 0.2 | 0.2 | 0.22 | 0.19 | 0.22 | |

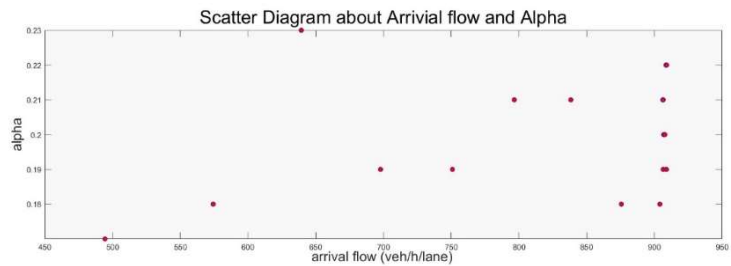

Figure 3 α-λ Scatter Diagram

In order to determine the correlation between arrival flow λ and weight coefficient α, the curve fitting on original data is needed. Using 6-order polynomial fitting, centering and scaling, it is found that fitting best, as shown in the following Figure 4.

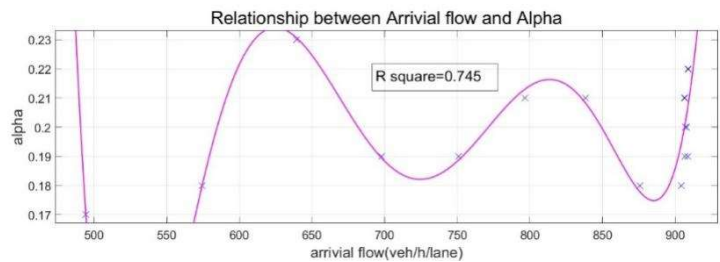

Figure 4 α-λ curve fitting diagram

By observing the above figure, we could find that the function curve of relationship between λ and α fluctuates up and down and the range is small. Furthermore, we could also find that with the intensity of sample point is lower, the range of fluctuation is gradually smaller. It can be inferred that when the number of sample point is infinite, function curve can be approximately represented by a horizontal line, indicating that the weight coefficient α is a constant regardless of how the arrival flow λ changes. Therefore, the value of α is a constant, due to the average of sample value is the unbiased estimation of mean value, the constant is the average value of each sample point's α. The formula is as follows

$$\alpha = \frac{\sum_{i=1}^{n} \alpha_i}{n}$$

where n is the number of samples; $\alpha_i$ is weight coefficient of the $i^{th}$ sample point.

Finally, α = 0.2.

In order to verify the accuracy of above process, calculating the discrete coefficient of each sample point weight coefficient is as follows

$$Vs = \frac{s}{x}$$

where s is the standard deviation of sample; s is average value of the sample.

After calculation, Vs is 0.085, this discrete coefficient is less than 0.1, indicating that the distribution of the weight coefficient α is relatively concentrated, which could verify above inference.

The value of weight coefficient α is 0.2, which shows clearly that the drivers in the closed lane change to the open lane in the merge area, among the factors that pushing driver to change lane, objective stimuli accounts for 20 %, and value stimuli accounts for 80 %. In other words, when drivers change lane in the merge area, the main reason for driving them to change lane is that the traffic condition of the closed lane in the merge area is getting worse, and the efficiency of driving is rapidly reduced, so as to pursue higher driving speed and driving efficiency.

In order to verify the accuracy of predicting queue length, the queue length of the CA model and the predicted one in our model are rounded under different arrival flow λ, so that the absolute error between above two is obtained.

Table 3 Absolute error under different arrival flow

| arrival flow(veh/h/lane) | 495 | 575 | 640 | 698 | 751 | 797 |
|---|---|---|---|---|---|---|
| absolute error(veh) | 0 | 0 | 0 | 1 | 0 | 1 |
| arrival flow(veh/h/lane) | 839 | 876 | 905 | 907 | 908 | 909 |
| absolute error(veh) | 1 | 2 | 3 | 0 | 1 | 1 |

From above Table 3 we could find that the absolute error between queue length of the CA model and the predicted one in our model is equal to or less than 1 vehicle under most of conditions, indicating that the prediction error of our model is not more than 1 vehicle under most of conditions. In summary, the prediction accuracy of our model is high, and this model is effective.

**4 Model Application**

About the model in this paper, we could use its prediction characteristic to generate reasonable solutions for the traffic bottleneck of work zone during the micro-traffic design stage. One of solutions is to determine the warning zone length according to different design traffic volumes, and the other is to use the predicted average queue length as a kind of evaluation indicator to select the best merge strategy.

4.1 Alert Length Determination

Warning zone length should not be too short, otherwise prone to congestion and collision. However, the warning zone should not be too long as well. Prolonging the warning zone requires bigger economic cost. Moreover, a too long warning zone makes a lot of vehicles change lane prematurely, resulting in low density in the closed lanes near the work zone, so that high density in the open lane has a negative impact on the utilization efficiency of closed lane and the road capacity of work zone. The appropriate warning zone length plays an important role in work zone optimization.

By the established estimation model, the warning zone length can be calibrated before the construction of work zone, and the appropriate warning zone length can be obtained. Under different design traffic flow V, the predicted value f (veh) of average queue length in our model is also changed by changing the value of the warning zone length L(m). The function f = f (L) of f to L is constructed. Considering the length of vehicle is generally less than or equal to 5 m, the value of L(m) is taken as multiple of 5. f decreases with the increase of L, and when the larger L is, the slower the value of f decreases, in other words, for any value of k greater than 1, there is always f (k-5) > f (k), and f (k-5) -f (k) > f (k) -f (k+5).

In this function, we could find the point of L = Lr. When the L<Lr, with the increase of L, f (L) decreases rapidly. Instead, when the L>Lr, f(L) decreases slowly. We can infer that L = Lr is a appropriate value of warning zone length, and the value of L at the inflection point is the appropriate warning zone length in our study. If f(k-5)-f(k)<0.5, it is clear that the downward trend is mild enough, it is inferred that the point whose value is L = k near the inflection point, so that the value of Lr is approximated to k. Finally, the results are showed in Figure 5 and Table 4.

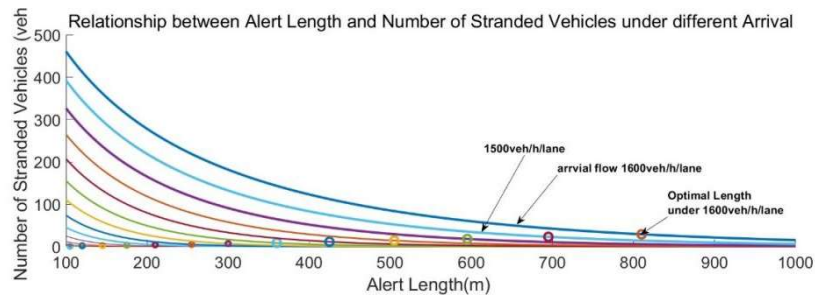

Figure 5 The relationship between average queue length and warning zone length

Table 4 The appropriate work zone length under different design volumes

| Design volume V(veh/h/lane) | 100~400 | 500 | 600 | 700 | 800 | 900 | 1000 |
|---|---|---|---|---|---|---|---|
| Appropriate warning zone length L(m) | 105 | 120 | 145 | 175 | 210 | 255 | 300 |

| Design volume V(veh/h/lane) | 1100 | 1200 | 1300 | 1400 | 1500 | 1600 |
|---|---|---|---|---|---|---|
| Appropriate warning zone length L(m) | 360 | 425 | 505 | 595 | 695 | 810 |

In the model validation part, we used the Meticulous R-STCA proposed by L. Fei et al. in the paper Analysis of traffic congestion induced by the work zone[33]. This CA itself has high simulation accuracy which is close to the real data. However, the appropriate merge length mentioned in this paper may have a little deficiency. As a sort of supplement, the warning zone and warning zone length mentioned in our study are the same as the merging area and merging length mentioned in reference, respectively. In this study, we use the warning zone and warning zone length.

L. Fei et al. give the appropriate warning zone length under different traffic flow densities. When the value of density is 0.14, according to the formula density=N/(2L-Lw), where N=868 (total vehicles within 10000s), the corresponding traffic flow is 156veh/h/lane, which is between 100veh/h/lane and 200veh/h/lane, and when the flow is 100veh/h/lane or 200veh/h/lane, according to our study, the appropriate warning zone length is 105m after linear interpolation. When the flow is 156veh/h/lane, the appropriate warning zone length is still 105 m. After that, the appropriate warning zone length (330m) given by L. Fei et al. under density is 0.14 compared with the appropriate warning zone length under the condition of density is 0.14 in our study. The flow of 156veh/h/lane corresponds to the vehicle generation probability (simulation parameters) p _ rate = 4.8%. Therefore, p _ rate is set to be 4.8 % in Meticulous R-STCA, and the value of the length of the warning zone is set to 330m and 105m respectively. The simulation time is 1000 tick (1 tick is equivalent to 1s in real condition), then average queue length (veh/h), average vehicle speed (km/h) and average travel time (s) are output. The output result is shown in table as follow Table 5.

Table 5 Results of two methods

| Warning zone length (m) | Average queue length (veh/h) | Average travel time (s) |
|---|---|---|
| 330 | 0.00 | 236.22 |
| 105 | 0.00 | 235.46 |

Comparing the method of warning zone length with the 330m one, the number of stranded vehicles in front of the work zone is 0.00 on average within an hour, and the difference between two methods about average travel time is only 0.76s, which is almost equal. The 105 m method on the length of the warning zone reduces 68% compared with the 330m one's, which significantly reduces the economic cost and improves the utilization rate of the closed lane. In conclusion, the 105m method proposed in our study is better.

4.2 Comparison and selection between difference merge strategies

Although our model could well predict queue length on the closed lane in work zone, the research background of our model is the traffic flow organization under normal merge strategy. In order to better solve the problem that too long queue length on the closed lanes in work zone exists under the existing traffic flow organization, it is necessary to optimize the corresponding merge strategy for traffic flow organization in the work zone. In this section, the existing merge strategies are introduced. After that, with the predicted average queue length as the evaluation index, the best merge strategy under different design traffic flow will be selected, finally the selection scheme of merge strategies in work zone under different design traffic flow is obtained.

At present, the existing merge strategies generally include normal merge strategy (NM), early merge strategy (EM) and improved signal-based merge strategy (IM), which will be introduced respectively below. NM is a kind of normal merge strategy, which means that no other management measures are taken except for the establishment of warning sign in front of a certain scale of the work zone, and the merging of vehicles is finished by self-organization. This method could be used in the case of low traffic flow, which helps to minimize the cost and reduce the complexity of driver's information processing. EM is the early merge strategy, which means that on the basis of the warning sign, in the long distance in front of the work zone, it is suggested that the vehicles on the closed lane change to the open lane, for achieving the purpose that all vehicles on the closed lane have completed the lane change process before arriving the work zone. Its root is to alleviate the contradiction between transportation supply and demand in front of the work zone by increasing the demand that vehicles change lane from close lane to open lane, so as to reduce the queue length on the close lane in front of work zone. Early merge strategy includes static early merge strategy and dynamic early merge strategy. The static early merge strategy refers to the location that giving drivers suggestion that changing lane from close lane to open lane is fixed, which reminds drivers to change lanes to open one in advance. Dynamic early merge strategy is a dynamic location where vehicles on close lane are suggested to change lane to open one. The limitation of early merge strategy is that although early merge strategy is believed to improve traffic safety, some researchers have found that it cannot increase throughput, especially in high traffic flow conditions [34-35]. If drivers merge into the target lane in advance and are delayed by the slower vehicle ahead, the travel time will increase [36]. In addition, no-passing zones are also regarded as a waste of space usage in the closed lane [27]. IM is a merge strategy based on intelligent dynamic adjustment traffic signal, which refers to the signal control through real-time computing of green time and dynamic conversion of signal stage. The open lane and the close lane are regarded as two approaches, and the signal control scheme of two-phase is similar to intersection's is adopted and realizing the specific signal timing scheme by determining the threshold and signal cycle. The advantage of IM is that it applies to a wider range of arrival traffic conditions without combining other merge strategies. However, IM merge strategy also has its limitations, a large number of radical mergers could be observed under high arrival traffic conditions exposes its security problems. Therefore, try not to use IM strategy to ensure the traffic safety.

In fact, the essence of EM is to change the transportation demand from the close lane to open lane. At a certain distance in front of the work zone, O increases significantly, and the value of d(x) increases accordingly, which close to s(x), in other words, more drivers choose to change lane in advance at position with more lane-changeable gaps. The essence of IM is to change transportation supply of the open lane. Taking the close lane as research subject, in the road of close lane before approach, the transportation supply of the open lane is s(x) under NM. Before the stop line of approach, due to the full line covering road does not allow vehicle to change lane, the transportation supply of open lane in such section is zero. When the traffic flow is high and if NM or EM is still used, s(x) will be almost zero at the downstream of the stop line. However, when the IM strategy is used, due to the control of traffic light, the transportation supply of open lane will increase significantly. According to the calculation formula of road capacity of signalized intersection, the transportation supply of open lane is at the downstream of stop line is $\frac{g}{C}$, which greatly improves the transportation supply. Where g is the accessible time of the vehicle on close lane, and C is cycle length.

The background of this model is in the work zone on close lane of two-lane highway under normal merge strategy. The average queue length under the appropriate warning zone length can be estimated, and such average queue length is called optimum average queue length. The result is as follows Table 6.

Table 6 optimal prediction queue length

| Arrival flow(veh/h/lane) | 200 | 300 | 400 | 500 | 600 | 700 | 800 |
|---|---|---|---|---|---|---|---|
| Appropriate warning zone length(m) | 105 | 105 | 105 | 120 | 145 | 175 | 210 |
| Optimum average queue length(veh) | 0 | 0 | 1 | 2 | 3 | 3 | 4 |
| Arrival flow(veh/h/lane) | 900 | 1000 | 1100 | 1200 | 1300 | 1400 | 1500 |
| Appropriate warning zone length(m) | 255 | 300 | 360 | 425 | 505 | 595 | 695 |
| Optimum average queue | 5 | 7 | 9 | 12 | 14 | 18 | 23 |

| length(veh) | | | | | | | |
|---|---|---|---|---|---|---|---|
| Arrival flow(veh/h/lane) | 1600 | 1700 | 1800 | 1900 | 2000 | 2100 | 2200 |
| Appropriate warning zone length(m) | 810 | 940 | 1080 | 1240 | 1410 | 1590 | 1775 |
| Optimum average queue length(veh) | 29 | 37 | 46 | 57 | 70 | 87 | 107 |

Above table describes the optimal queue length corresponding to different arrival flow under NM, optimal queue length not only satisfying the increase of the length of warning zone but also ensuring queue length is lower than most of queue length corresponding to shorter alert length. So it is the best queue length under NM.

Although our model could obtain the prediction queue length under NM strategy, the queue length is still long when the arrival flow is high due to the existing merge strategy. Therefore, different merge strategy should be selected under different traffic flow and queue length. Queue length and arrival flow are king of good evaluating indicators, because they directly reflect level of service (LOS) with certain strategy. Therefore, it is necessary to select different merge strategies according to different queue length or arrival flow, due to arrival flow is common indicator that variable model will use, so that according to the threshold of average queue length for different merge strategies determining threshold of arrival flow in our model is crucial. In this section, the threshold of average queue length is initially selected, which corresponds to arrival flow, then such threshold is verified by the existing verification dataset in another research. Finally, the optimal merge strategy under different optimal queue length is determined according to the arrival flow threshold. Specific process is as follow

When the average queue length(veh/h) is zero in our model, no matter what kind of merge strategy is adopted, LOS of road is basically consistent. Therefore, selecting the lowest cost of merge strategy could maximize effect, and NM strategy could achieve this. Overall, when the average queue length is zero, NM strategy is selected.

When the average queue length is greater than 0, NM strategy isn't suitable for the smooth operation of traffic flow organization. Therefore, the optimal strategy will be selected between EM and IM. It can be seen from the above introduction of the merge strategy that the traffic safety of IM is relatively low. Under the same traffic condition, EM has higher priority. From the perspective of transportation supply and demand, if EM strategy is adopted, the willingness of vehicle change lane to the open lane will increase, that is, the transportation demand will increase. When the transportation is not lacking, increasing the demand of changing lane on the close lane can help vehicles on the close lane merge into open lanes more quickly, thus slowing down the congestion pressure of close lane and reducing the average queue length in the close lane.

When the average queue length is greater than a certain value, the queue length of the close lane will increase significantly under NM strategy. If EM strategy is adopted, the transportation supply of the open lane could not meet the transportation demand of the close lane, that is, the gap in the open lane is too small. Therefore, congestion only be avoided by increasing transportation supply. IM strategy could effectively increase the transportation supply of open lane, so that the demand in close lane is met and the average queue length on close lanes could be reduced.

About the method that selection of threshold value, we could determine a certain value of average queue length, when the average queue length is greater than it, the subsequent queue length will increase significantly with the increase of arrival flow. The specific process are as follows.

The model will be input a large arrival flow, and the value of this example is 5000veh/h/lane. Although this value does not exist in real condition, it is very helpful for the selection of queue length threshold. Then difference of average queue could be obtained by the optimal queue length under higher arrival flow is subtracted from the optimal queue length under lower arrival flow in interval of 100veh/h/lane (from 200veh/h/lane to 5000veh/h/lane, 48 groups in total). The following bar chart Figure 6 could be obtained according to the difference data.

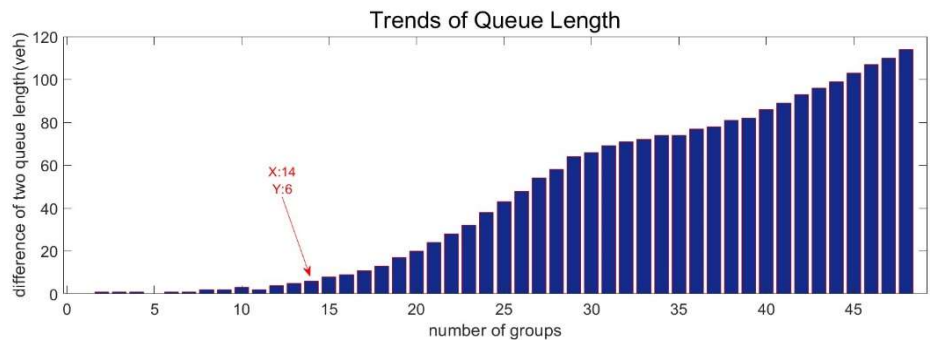

Figure 6 Trend about the difference of average queue length in our model

From the above statistical graph, when the difference of the average queue length is 6, the average queue length will increase significantly with the increase of number of groups. This difference of average queue length corresponding to NM strategy average queue length is greater than or equal to 37.

To sum up, when the average queue length is 0, NM strategy should be adopted; when the average queue length greater than 0 and less than 37, EM strategy should be adopted; when the best queue length greater than and equal to 37, IM strategy should be adopted.

Finally, in our model, when the design traffic flow is less than 400veh/h/lane, the merge strategy should be NM; when the flow is greater than or equal to 400veh/h/lane and less than 1700veh/h/lane, the merge strategy should adopt EM; when the flow is greater than or equal to 1700veh/h/lane, the merge strategy should adopt IM.

The results of strategy choice are shown in Table 7.

Table 7 The results of strategy choice

| arrival flow(veh/h/lane) | 200 | 300 | 400 | 500 | 600 | 700 | 800 | **NM** |
|---|---|---|---|---|---|---|---|---|
| optimum len_alert(m) | 105 | 105 | 105 | 120 | 145 | 175 | 210 | **EM** |
| optimum queue length(veh) | 0 | 0 | 1 | 2 | 3 | 3 | 4 | **IM** |
| arrival flow(veh/h/lane) | 900 | 1000 | 1100 | 1200 | 1300 | 1400 | 1500 | |
| optimum len_alert(m) | 255 | 300 | 360 | 425 | 505 | 595 | 695 | |
| optimum queue length(veh) | 5 | 7 | 9 | 12 | 14 | 18 | 23 | |
| arrival flow(veh/h/lane) | 1600 | 1700 | 1800 | 1900 | 2000 | 2100 | 2200 | |
| optimum len_alert(m) | 810 | 940 | 1080 | 1240 | 1410 | 1590 | 1775 | |
| optimum queue length(veh) | 29 | 37 | 46 | 57 | 70 | 87 | 107 | |

The verification dataset of another research will be used to verify the above conclusions, and data are shown in Table 8.

Table 8 maximum queue length under NM, EM and IM strategy

| arrival flow (veh/h/lane) | 500 | 600 | 700 | 800 | 900 | 1000 | 1100 | 1200 | 1300 |
|---|---|---|---|---|---|---|---|---|---|
| NM | 2 | 4 | 5 | 6 | 10 | 14 | 19 | 28 | 32 |
| EM | 3 | 2 | 4 | 4 | 8 | 8 | 12 | 16 | 21 |
| IM | 3 | 4 | 4 | 4 | 8 | 11 | 12 | 16 | 25 |
| arrival flow (veh/h/lane) | 1400 | 1500 | 1600 | 1700 | 1800 | 1900 | 2000 | 2100 | 2200 |
| NM | 48 | 57 | 65 | 74 | 86 | 96 | 100 | 107 | 108 |
| EM | 32 | 41 | 52 | 59 | 76 | 82 | 87 | 95 | 103 |
| IM | 34 | 40 | 44 | 53 | 52 | 53 | 55 | 56 | 57 |

According to above table, when the maximum queue length is greater than 0 and not large significantly, the difference between the maximum queue length of EM strategy and IM strategy in the close lane is small, and the maximum queue length of EM in the open lane is significantly smaller than that of IM. Therefore, when the maximum queue length is greater than 0 and not large significantly, EM strategy is selected. When the maximum queue length is large, the gap between the maximum queue length of the EM strategy and the IM strategy on the close lane becomes larger. Although the queue length in open lane under EM strategy is smaller than the IM strategy's, the maximum queue length of IM on the close lane is significantly smaller than that of EM, and the throughput under IM is larger than that of EM. Therefore, when the maximum queue length is high significantly, the IM strategy is selected. Regarding the critical value between not large significantly and large significantly, the maximum queue length difference between IM strategy and EM strategy is selected as the indicator. When the difference begins to increase sharply after a certain value, such value is the critical value. The specific bar graph is as follows Figure 7:

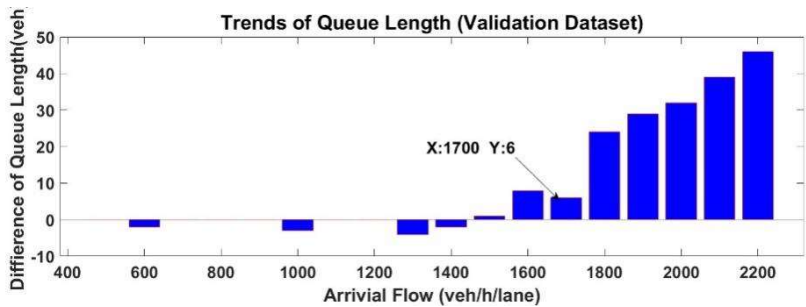

Figure 7 Trends about difference of maximum queue length between EM and IM（Validation Dataset）

From above figure, when the maximum queue length difference between EM strategy and IM strategy is 6, the corresponding arrival flow is 1700veh/h/lane. After that, with the increase of arrival flow, the maximum queue length difference will increase rapidly. The maximum queue length corresponding to NM strategy is greater than 32. In summary, when the arrival flow in the verification data is greater than 300veh/h/lane and less than 1800veh/h/lane, EM strategy should be adopted. When the arrival flow is greater than or equal to 1800veh/h/lane, IM strategy should be adopted.

Above two conclusions about our model and previous research are applied to common different traffic flow, and the strategy selection results are basically consistent. The comparison Table 9 of the two conclusions is as follows.

Table 9 comparison between two conclusions

| arrivial flow(veh/h/lane) | 200 | 300 | 400 | 500 | 600 | 700 | 800 |
|---|---|---|---|---|---|---|---|
| merge strategy_OURS | NM | NM | EM | EM | EM | EM | EM |
| Merge strategy_VALIDATION | NM | NM | EM | EM | EM | EM | EM |
| arrivial flow(veh/h/lane) | 900 | 1000 | 1100 | 1200 | 1300 | 1400 | 1500 |
| merge strategy_OURS | EM | EM | EM | EM | EM | EM | EM |
| merge strategy_VALIDATION | EM | EM | EM | EM | EM | EM | EM |
| arrivial flow(veh/h/lane) | 1600 | 1700 | 1800 | 1900 | 2000 | 2100 | 2200 |
| merge strategy_OURS | EM | IM | IM | IM | IM | IM | IM |
| merge strategy_VALIDATION | EM | EM | IM | IM | IM | IM | IM |

It can be seen from above Table 8 that the conclusion of our model and the verification conclusion are different only when the arrival flow is 1700veh/h/lane, indicating that the results of our model have certain credibility in the comparison and selection of merge strategy. 1700veh/h/lane is the threshold value selected in merge strategy comparison. When the traffic flow is 1700veh/h/lane, selecting IM is benefit to balancing queue length between two lanes, maximizing the utilization rate of road resources, besides, it will more actively respond to the uncertainty of arrival flow in real condition.

**5 Summary and outlook**

In this study, an estimation model for the average queue length at work zone is established. In the process of establishing the model, this study fully takes into account the distribution of the gaps in the open lane and driver 's lane-changing intention in closed lane, as well as the weight of the two lane-changing driving factors. The estimation model is verified based on cellular automata. Finally, the estimation model is used to calibrate the appropriate warning zone length, and select the appropriate merge strategy for work zone.

However, there are still some deficiencies in this study. First, this study only discusses the case of 2to1 two-lane highway work zone, and more cases can be discussed later. Second, when verifying the model in this study, the repeating number of simulations is 10, and the amount of data is not large, so the repeating number of simulations can be increased in future studies. Third, in this study, while selecting the work zone merge strategies, only NM, EM, IM strategies are discussed, and other lane change strategies can be joined later.